\documentclass[manuscript]{acmart}

\usepackage{bm}
\usepackage{bbm}
\usepackage{booktabs}
\usepackage{hyperref}
\usepackage{dirtytalk}
\usepackage{graphicx}
\graphicspath{ {./figures/} }
\usepackage{xifthen}
\usepackage[USenglish]{babel}
\usepackage{xspace}
\usepackage{xcolor}
\usepackage{mathtools}
\usepackage{cleveref}
\usepackage{subcaption}
\usepackage{dirtytalk}
\usepackage[ruled,vlined]{algorithm2e}
\usepackage{makecell}
\usepackage{enumitem}

\crefname{equation}{Eq.}{Eq.}
\crefname{section}{Section}{Sections}
\crefname{subsection}{Section}{Sections}
\crefname{subsubsection}{Section}{Sections}
\crefname{figure}{Figure}{Figures}
\crefname{table}{Table}{Tables}
\crefname{algocf}{Algorithm}{Algorithms}

\usepackage{pifont}
\newcommand{\cmark}{\ding{51}}
\newcommand{\xmark}{\ding{55}}

\newcommand{\xii}{\bm{x}_{i}}
\newcommand{\xuu}{\bm{x}_{u}}
\newcommand{\xiitest}{\bm{\bar{x}}_{i}}
\newcommand{\xuutest}{\bm{\bar{x}}_{u}}
\newcommand{\pinv}[2]{\ifthenelse{\isempty{#1}}{p_{#2}^{-1}}{p_{#1,#2}^{-1}}}
\newcommand{\p}[2]{\ifthenelse{\isempty{#1}}{p_{#2}}{p_{#1,#2}}}

\newtheorem{thm}{Theorem}[section]
\newtheorem{definition}[thm]{Definition}
\newtheorem{subdefinition}{Definition}[thm]

\theoremstyle{definition}

\AtBeginDocument{%
  \providecommand\BibTeX{{%
    \normalfont B\kern-0.5em{\scshape i\kern-0.25em b}\kern-0.8em\TeX}}}

\setcopyright{acmcopyright}
\copyrightyear{2021}
\acmYear{2021}

\acmConference[]{}{}{}
\acmBooktitle{}
\acmPrice{}
\acmISBN{}

\begin{document}

\title{Situating Recommender Systems in Practice: \\Towards Inductive Learning and Incremental Updates}
\renewcommand{\shorttitle}{Towards Inductive Learning and Incremental Updates}
\author{Tobias Schnabel}
\affiliation{%
  \institution{Microsoft}
  \city{Redmond} 
  \state{WA}
  \country{USA}}
\email{Tobias.Schnabel@microsoft.com}\authornote{equal contributions}

\author{Mengting Wan}
\affiliation{%
  \institution{Microsoft}
  \city{Redmond} 
  \state{WA}
  \country{USA}}
\email{Mengting.Wan@microsoft.com}\authornotemark[1]

\author{Longqi Yang}
\affiliation{%
  \institution{Microsoft}
  \city{Redmond} 
  \state{WA}
  \country{USA}}
\email{Longqi.Yang@microsoft.com}\authornotemark[1]


\begin{abstract}
  With information systems becoming larger scale, recommendation systems are a topic of growing interest in machine learning research and industry. Even though
  progress on improving model design has been rapid in research, we argue that many advances fail to translate into practice because of two limiting assumptions. 
  First, most approaches focus on a \emph{transductive learning} setting which cannot handle unseen users or items and second, many existing methods are developed for static settings that cannot incorporate new data as it becomes available. We argue that these are largely impractical assumptions on real-world platforms where new user interactions happen in real time. In this survey paper, we formalize both concepts and contextualize  
  recommender systems work from the last six years. We then discuss why and how future work should move towards \textbf{inductive learning} and \textbf{incremental updates} for recommendation model design and evaluation. In addition, we present best practices and fundamental open challenges for future research.
\end{abstract}

\ccsdesc[500]{Information systems~Information retrieval}
\ccsdesc[500]{Information systems~Personalization}
\ccsdesc[500]{Information systems~Recommender systems}

\keywords{Recommender system; inductive learning; incremental updates }


\maketitle


\section{Introduction}
Recent years have witnessed a dramatic increase in research on recommendation systems. 
Today's recommenders have gone beyond classic matrix factorization~\cite{koren2009matrix, hu2008collaborative} and incorporated many state-of-the-art machine learning techniques, such as recurrent neural networks~\cite{hidasi2015session,hidasi2018recurrent}, convolutional neural networks~\cite{he2016vbpr, tang2018personalized}, graph neural network~\cite{ying2018graph, he2020lightgcn, zhang2019inductive, wang2019neural}, reinforcement learning~\cite{wang2020kerl, zheng2018drn, chen2019top}, and meta-learning~\cite{zhang2020retrain, vartak2017meta}. In addition, the data used to fuel recommenders has become much richer in nature, e.g., through search queries~\cite{zamani2020learning}, images~\cite{he2016vbpr, kang2017visually, hsieh2017collaborative}, contextual information~\cite{hansen2020contextual, aliannejadi2021context}, knowledge graphs~\cite{chen2020jointly, xu2020product, wang2020make}, or reviews in natural language~\cite{wang2018explainable, zheng2017joint, sun2020dual}. 

However, despite the rapid research progress over the past years, industry practitioners typically find it hard to apply many state-of-the-art models to real-world settings due to a fundamental mismatch between model assumptions and the reality of large-scale production platforms. 
In this paper, we address two core assumptions which we believe are most responsible for this disconnect.

First, most of the recommendation models developed to date focus purely on a \textit{transductive learning} setting which assumes that all testing users or items are known during training (\cref{sec:definition-inductive}). 
However, in practice, it is virtually impossible to know all of the users and items beforehand when training recommendation models because users and items come and go continuously in real time~\cite{ma2020temporal,covington2016deep,smith2017two,tsagkias2020challenges}. 
In addition, the cost of training recommender systems on the entirety of logged interaction records becomes increasingly burdensome since many of today's platforms have already reached multi-billion or trillion interaction counts (e.g., on Twitter~\cite{vartak2017meta}, Youtube~\cite{covington2016deep}, Pinterest~\cite{ying2018graph}, Alibaba~\cite{wang2018billion}). 
The observations above motivate
our first call for increased investment into \textbf{inductive models} 
which support unseen users and items during testing as well as support training on subsamples of the entire data, offering a straightforward way to trade-off compute time with model accuracy on large datasets.

Second, current literature often assumes a \textit{static} recommendation setting where user preferences or item contexts remain unchanged over time. This assumption is generally inconsistent with real-world recommendation scenarios where users interact with the system continuously and can leave or enter at any given time~\cite{koren2009collaborative,amatriain2015recommender,covington2016deep, chen2019top,ma2020temporal,smith2017two}. To accommodate these dynamics, real-world systems should be able to provide \textit{fresh} recommendations, otherwise they may substantially harm user experiences and business goals, e.g., by recommending products that people have already bought and no longer have the need for, or by not being responsive enough to already downvoted content. 
A common strategy that practitioners often employ is to refresh models at regular intervals~\cite{chen2019top}.
As this is expensive, trained models are only refreshed every couple of days or weeks, causing stale or obsolete recommendations. 
We therefore call for models that support efficient \textbf{incremental updates} as new data comes in. This can be accomplished in complementary ways (\cref{sec:definition-incremental}) 
-- (i) through incremental re-training which updates model parameters efficiently; or (ii) through incremental inference which induces new predictions efficiently while keeping the model parameters fixed. 

In summary, we argue that future research should more deeply invest into recommendation models that support \textbf{inductive learning} and \textbf{incremental updates}, so that (i) models can generalize to unseen users and items and (ii) can efficiently update recommendations with new interaction records. This paper discusses the state of the research based on extensive evidence from 223 related papers published over the past six years complemented with case studies on specific model types. We conclude with suggestions on best practices as well as with a roadmap of future research directions.

\subsection{Related Work}

It is worth noting that many of the recommender systems that implicitly support incremental updates and indutive learning were developed in industry labs \cite{vartak2017meta,covington2016deep,ying2018graph,pi2019practice,wang2018billion}. For example, items were represented using their explicit features in the Google Play Store app recommendation model \cite{cheng2016wide} and the Alibaba advertising click-through rate prediction model \cite{pi2019practice} so that these systems can naturally support unseen items during serving (i.e., item-inductive learning). In addition, users were represented by aggregating over the items they consumed in the Youtube video recommender systems \cite{covington2016deep,chen2019top} so that these models can be easily applied on unseen users (i.e., user-inductive learning). The system of LinkedIn feed recommendations was purposely designed to support incremental inference and affordable re-training, where item features, user features as well as their affinity features (e.g., the past interactions between the user and the post author) were carefully engineered and precomputed  \cite{agarwal2014activity, agarwal2015personalizing}. However, as our survey below will show, these papers are still in the minority and lack a formal framework for categorization which we provide in this paper.

There are several related surveys in the area of recommender systems, which typically focus on a particular type of recommendation scenarios or a group of fast-evolving technologies applied in recommender systems. Sequence-aware recommendation (including session-based recommendation) refers to a group of scenarios where item recommendations to a user are made based on a temporally ordered list of past user actions \cite{quadrana2018sequence, wang2021survey}. These recommendation scenarios require users' temporal dynamics being considered; therefore user-inductive learning and incremental updates become critical. Another line of related surveys overview the application of reinforcement learning and active learning techniques in recommender systems \cite{chen2021survey,afsar2021reinforcement,elahi2016survey,rubens2015active}. These techniques are set to address the dynamic environments by adapting the model based on users' real-time feedback, and thus closely related to the concept of incremental updates. 

In this paper, we formalize the two key requirements---inductive learning and incremental updates---in the real-world industrial recommender systems. Our work differs from existing surveys by raising these two primary mismatches in model assumptions. 
By doing so, we hope to call for attentions on inductive and incremental recommendation, and to invite future research to fundamentally address these challenges beyond ad hoc industry practices.

\section{Key concepts and definitions}\label{sec:definition}

Throughout this paper, we assume the common recommendation setting of predicting items $i$ to users $u$. 
We represent knowledge about users and items through context vectors:
\begin{itemize}
    \item User context $\xuu \in \mathcal{U}$ encodes all relevant information about a user; often this contains a unique user identifier.
    \item Information context $\xii \in \mathcal{I}$ represents all relevant item information; again, this may simply be an unique item identifier.
\end{itemize}
During learning we are given a training set for each user, 
$\mathcal{D}_{\mathit{train}}(u) = \{(\xuu, \xii, r^*_{u,i})\}$, 
and we want to learn a model that assigns utility scores $r_{u,i} \in \mathbb{R}$ by optimizing a loss function
\begin{equation}
    \mathscr{L}_u\big(\{r_{u,i}\}, \{r^*_{u,i}\}\big), \label{eq:task}
\end{equation}
so that the model fares well on a test set $\mathcal{D}_{\mathit{test}}(u) = \{(\xuutest, \xiitest, r^*_{u,i})\}$. 
In traditional implicit matrix factorization, the user context corresponds to the user identifier, $\xuu = [u]$, and the item context is mapped to the item identifier, $\xii = [i]$. Interacted user-item pairs receive $r^*_{u,i} = 1$, all other pairs are assigned zero. The test set typically consists of one held-out item per user and the training set contains all remaining interacted user-item pairs.

In the task definition above \cref{eq:task}, 
the loss function $\mathscr{L}$ can be instantiated in a variety of ways -- e.g., item-wise as is commonly done in classification-based and regression-based recommender systems, or listwise for top-$N$ recommendation~\cite{shi2010list}. 

The individual scores are typically represented through a scoring function that maps user-item contexts to a real-valued score:
\begin{equation}
    h: \mathcal{U} \times \mathcal{I} \rightarrow \mathbb{R}: (\xuu, \xii) \mapsto r_{u,i}~ .
    \label{eq:model}
\end{equation}
To make use of information across users, typically a single model is estimated to do well on \emph{all} users in the test set simultaneously, $\mathcal{D}_{\mathit{test}} = \bigcup_{u} \mathcal{D}_{\mathit{test}}(u)$, averaging or summing the individual per-user performances $\mathscr{L}_u$. 
Let $\mathcal{I}_{\mathit{train}}$ denote all item contexts seen in the training set and $\mathcal{I}_{\mathit{test}}$ all item contexts seen in the test set. $\mathcal{U}_{\mathit{train}}$ and $\mathcal{U}_{\mathit{test}}$ are defined analogously for the user contexts.
Also, let $\theta$ 
denote the current parameters (for a parametric model) or representation (for a non-parametric model) of function $h$. 

We are now ready to introduce the key concepts of this paper. 

\subsection{Inductive vs. transductive learning} \label{sec:definition-inductive}
We start with the most common setting in recommendation -- transductive learning where all user and item contexts in which the algorithm will be tested on are available during training. In both scenarios, we are given a training set 
$\mathcal{D}_{\mathit{train}} = \{(\xuu, \xii, r^*_{u,i})\}$ 
which is the union of all individual training sets for users $\mathcal{D}_{\mathit{train}}(u)$. Transductive and inductive learning now differ in what information about the test set we have at training time.

\begin{definition}[Transductive learning]
In transductive learning, the algorithm knows all item contexts $\mathcal{I}_{\mathit{test}}$ as well as user contexts $\mathcal{U}_{\mathit{test}}$ that will occur in the test set upfront. Put differently, the algorithm will not be tested on out-of-domain contexts.
\end{definition}

In most recommendation settings, the sets $\mathcal{U}_{\mathit{test}}$  and $\mathcal{I}_{\mathit{test}}$ are implicitly defined at training time to be the same as the ones seen in training. Returning to our earlier example, matrix factorization is transductive because at training time, we implicitly assume that all test contexts will be the same as during training, $\mathcal{I}_{\mathit{test}} = \mathcal{I}_{\mathit{train}}$  and 
 $\mathcal{U}_{\mathit{test}} = \mathcal{U}_{\mathit{train}}$.

\begin{definition}[Inductive learning]
In inductive learning, the algorithm has no information about the test set at training time. Hence, the algorithm must be able to work on \underline{unseen} 
user-item contexts during testing.
\end{definition}

There are two hybrid scenarios that arise when either information about user contexts or information about item contexts in the test set is available for training.

\begin{subdefinition}[User-inductive learning]
In user-inductive learning, the algorithm knows only the item contexts $\mathcal{I}_{\mathit{test}}$ that will occur in the test set at training time. This requires the recommender to be able to accommodate unseen user contexts during testing. 
\end{subdefinition}

\begin{subdefinition}[Item-inductive learning]
In item-inductive learning, the algorithm knows only the user contexts $\mathcal{U}_{\mathit{test}}$ that will occur in the test set at training time. 
\end{subdefinition}


We note that complementary to the learning scenarios above is the so-called \emph{cold start problem}~\cite{maltz1995pointing,schein2002methods}. In cold start recommendation, the focus lies on model performance behaves when not enough information about users or items is available (either at test or training time). This is independent of whether the model supports transductive or inductive inference.

\subsection{Incremental updates} \label{sec:definition-incremental}
We now turn to the computational requirements of incorporating new data into recommender models. 
Ideally, we would like a recommendation algorithm to update its predictions rapidly as new data or information becomes available. This is particularly important when data arrives in a \emph{streaming} fashion, i.e., data arrives at discrete timesteps $t=1, \ldots, T$ inducing an ordering of data points $(\xuu^{(t)}, \xii^{(t)}, \cdot)$.
There are two main pathways for incorporating new data; one can add the new data to the training set and re-train the model, or add it to the test set and re-run the inference step. 

\begin{definition}[Incremental inference.]
We say that an algorithm affords incremental inference if the cost of inferring new recommendations when adding a single data point to a users test set $\mathcal{D}_{\mathit{test}}(u)$ does neither depend on the size of the train set $|\mathcal{D}_{\mathit{train}}|$ or the size of test set $|\mathcal{D}_{\mathit{test}}(u)|$. This means that an algorithm cannot go through all datapoints in the test set of a user or the training set again.
\end{definition}

Note that the definition above does not specify how an algorithm needs to implement this -- it can be as simple as caching previous model outputs or might require sophisticated data structures.

The other update one can make to a recommender model is changing its training set $\mathcal{D}_{\mathit{train}}(u)$. This motivates that a recommender model can quickly process changes to its training set. 

\begin{definition}[Incremental re-training.]
    We say that an algorithm affords incremental re-training if the cost of updating $\theta$ to accommodate a new data point $(\xuu^{(t+1)}, \xii^{(t+1)},  r_{u,i}^{*(t+1)})$ does not depend on the the size of the training set $|\mathcal{D}_{\mathit{train}}|$. In particular, this means that an algorithm cannot go through all previous data points to update $\theta$.
\end{definition}

There are a number of related concepts that connect to incremental updates, but are worth discussing to emphasize their distinctiveness.
First, streaming recommendation and incremental updates critically differ on a conceptual level: incremental updates (whether it is through incremental inference or incremental re-training) are a \emph{model property} whereas streaming recommendation is a \emph{data property}. This is an important distinction to make since it has implications on how models can be used in practice, e.g., on how model parameters need to be stored.
A second related concept is cold start recommendation which focuses on model performance under varying amounts of training data, i.e., data efficiency. This is orthogonal to whether a model supports incremental updates – for example, it is possible to look at cold start performance of a transductive model, retrained at every timestep. 

Even though \emph{online machine learning} is a related concept, it only refers to the overarching setting of learning under sequential data streams. One important difference is that it does not posit any computational constraints on the updates -- for example, the popular follow-the-leader approach requires full re-training with each added data point~\cite{shalev2011online}, and is thus not incremental. Another difference is that so far, we have tacitly assumed stationarity -- namely that the data comes from an i.i.d. distribution -- which online algorithms do not generally require. We will re-visit the issue of stationarity in our discussion. Another related concept in machine learning is \emph{fine-tuning} (also known as \emph{warm starting}) which is model re-training with the initial parameters set to the previous solution. Even though fine-tuning has been observed in practice to accelerate re-training by speeding up convergence~\cite{rendle2008online}, it does not give any guarantees with respect to the computational costs involved.



\section{State of research} \label{sec:state-of-research}

To better understand the current state of research in the recommendation space regarding inductive learning and incremental updates, we conducted a systematic literature review of published work from three representative conference venues over the last six years. Our inclusion criteria were as follows. Work had to be published at either SIGIR, WSDM or TheWebConf (formerly known as WWW) between 2016 and 2021 as a full conference paper. We only considered papers that proposed a new recommendation algorithm and evaluated it empirically on at least one  recommendation 
dataset. 
Only few papers that mentioned recommendation did not meet these criteria. 
After filtering, we were left with 88 relevant papers from SIGIR, 72 papers from WSDM, and 63 papers from TheWebConf.

We then annotated according to whether it would require a practitioner to make any changes to the model code in order to achieve a certain property. Our annotation dimensions were:

\begin{description}
    \item[Generalization scenario (modeling).] Fully inductive if the approach used only features for both users and items or it was explicitly pointed out that they were fully inductive. For example, \citet{zheng2017joint} represents items through their review texts and users through the reviews they have written. Another example is the mixture of experts approach by \citet{xie2021real} that maps all user and item features to deeper feature spaces. We labeled papers as user-inductive (item-inductive) if only user features (item-features) were used or it the paper explicitly showed how to perform inductive inference. All other papers were labeled as transductive.
    \item[Generalization scenario (evaluation).] Here, we applied a stricter notion as we only considered the experimental setup in which approaches were tested. This means that even though a model theoretically could support inductive inference, it would not be counted here if this capability hadn't been evaluated empirically.
    \item[Sequential modeling.] Annotated with \say{yes} if approaches considered time information in data points (e.g., session-based recommender systems).
    \item[Incremental re-training.] Annotated with \say{yes} if the paper explained explicitly how to perform incremental re-training.
    \item[Incremental inference.]  Annotated with \say{yes} if the paper explained explicitly how to perform incremental inference (e.g., through memorization or caching).
\end{description}

\begin{figure}
	\centering
	\begin{subfigure}[b]{0.45\linewidth}
\centering
\includegraphics[width=\textwidth]{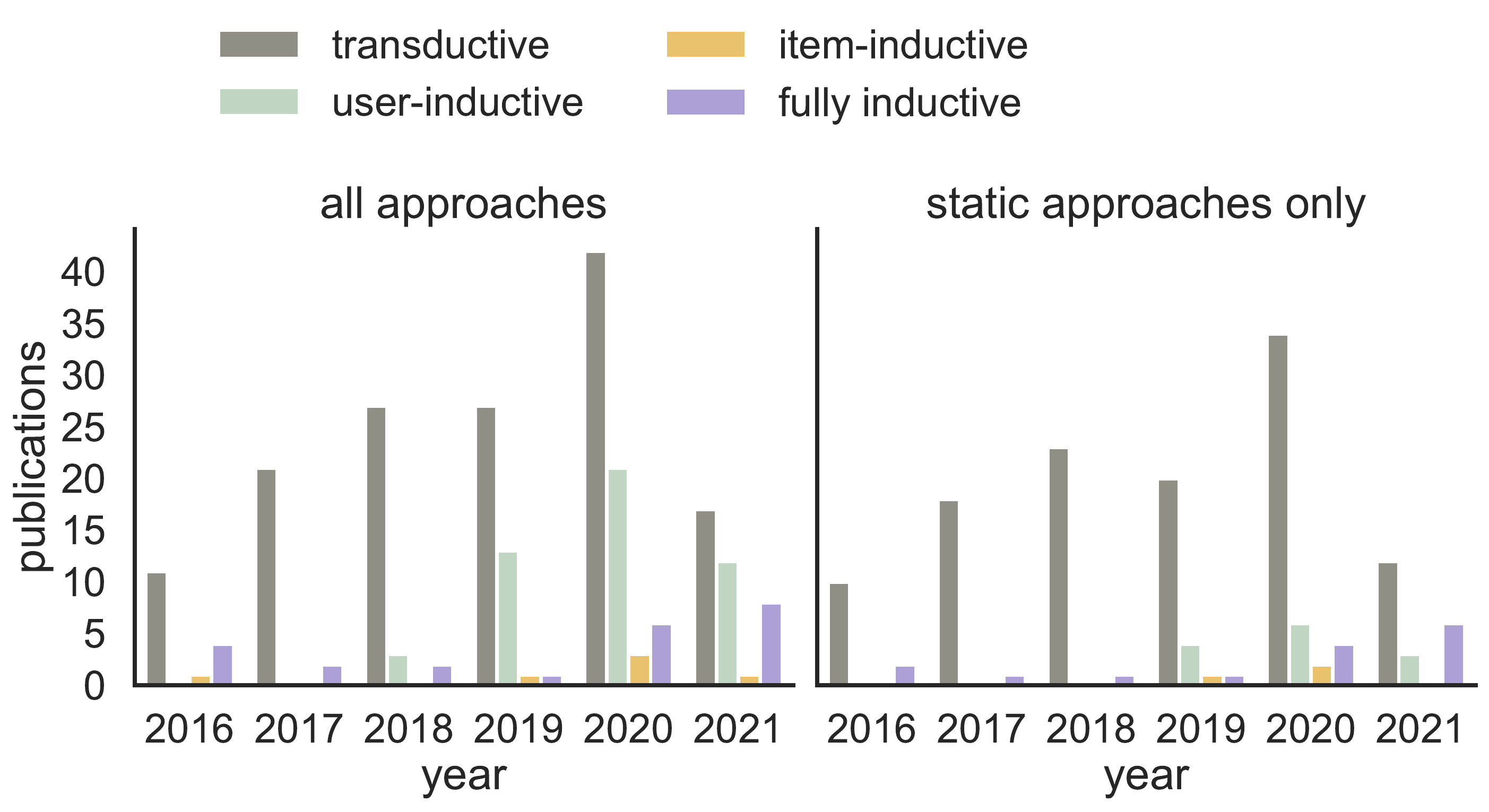}
\caption{Generalization scenarios supported by all models (left) or only static models (right). The overwhelming majority of models only support a transductive setting.}
\label{fig:generalization_learn}
	\end{subfigure}%
\qquad
\centering
	\begin{subfigure}[b]{0.5\linewidth}
\centering
\includegraphics[width=\textwidth]{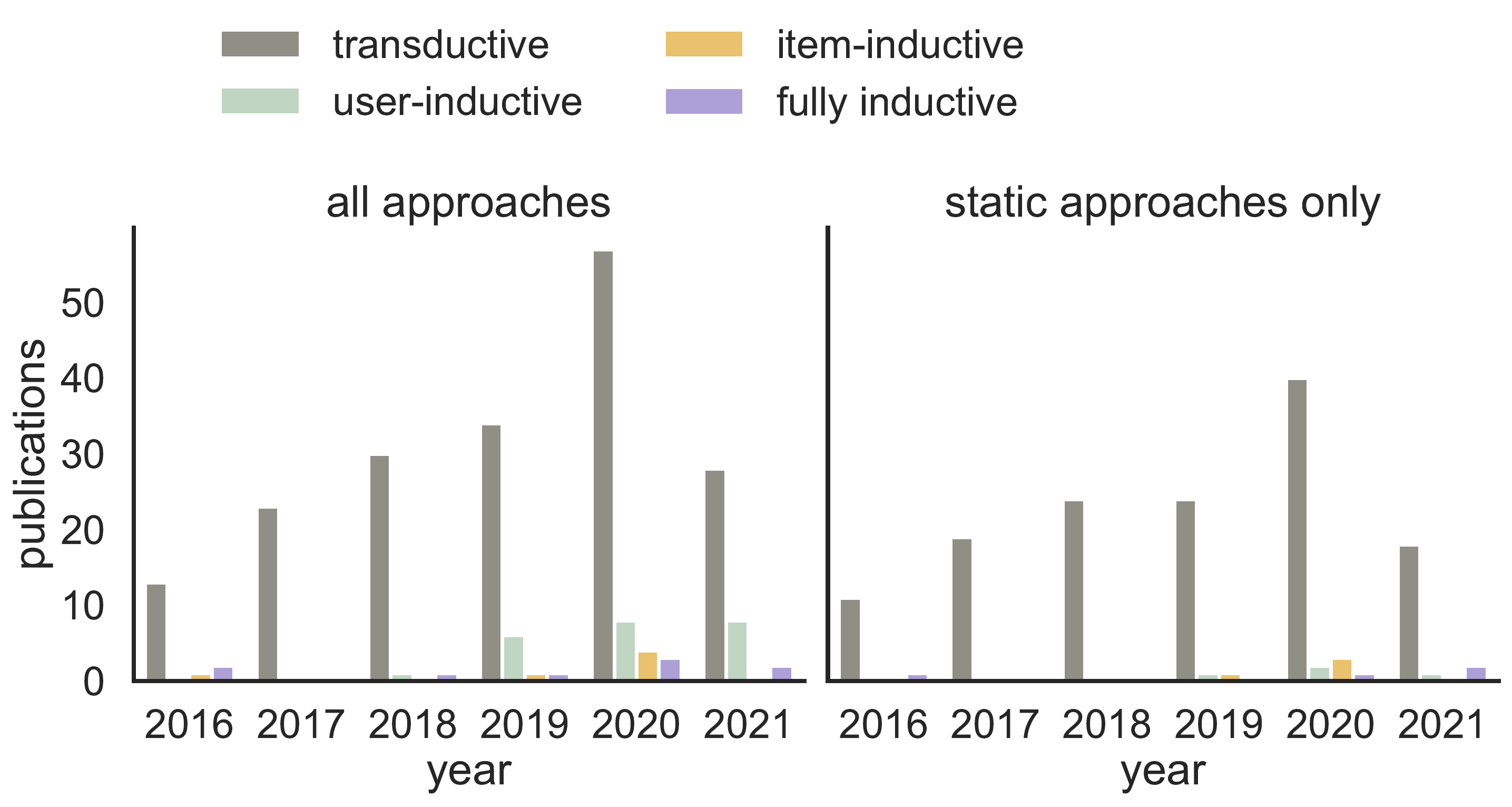}
\caption{Generalization scenarios that approaches were actually evaluated on. Left are all approaches, right are static approaches only. Again, most evaluation only was done in a transductive setting.}
\label{fig:generalization_eval}
	\end{subfigure}%
	
    \caption{Generalization scenarios targeted by recent work.}

\end{figure}



\subsection{Generalization scenarios}
As we can see from \cref{fig:generalization_learn} (left), most approaches only support transductive scenarios. Least frequent were approaches supporting item-inductive learning, followed by approaches that work in a fully inductive setting. Interestingly, the number of approaches supporting user-inductive  settings grew considerably in the last three years at a higher rater than any other category. However, this is mostly due to increased interest in sequential settings as the right-hand plot in \cref{fig:generalization_learn} shows in which we filtered out sequential models. On a side note, the overall number of submissions in the recommendation space has grown over the past six years, demonstrating the increasing importance and relevance of recommender systems.

Moving on to evaluation setups plotted in the \cref{fig:generalization_eval}, 
we can see that there are even fewer approaches that were empirically evaluated in the respective generalization scenarios. This means that there are a substantial portion of approaches that never are evaluated in an (partially) inductive scenario even though they do support it. We believe that this is due to a lack of established baselines and experimental protocols for such scenarios and will discuss more details and possible solutions in \cref{sec:guidance}. 

\begin{figure}[tbp]
\begin{subfigure}{0.4\textwidth}
\includegraphics[width=0.9\linewidth]{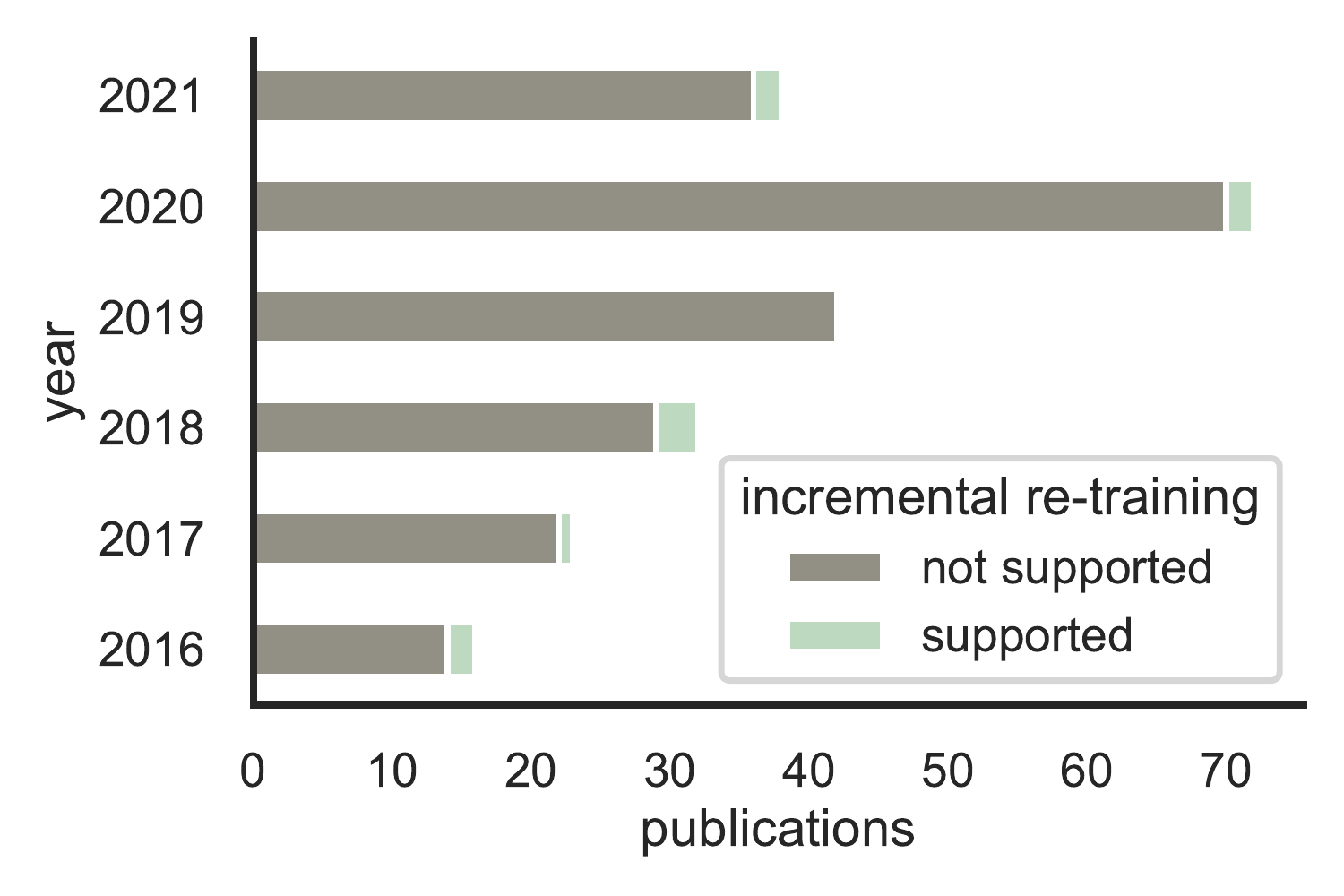}
\caption{Incremental re-training.}
\label{fig:retraining}
\end{subfigure}
\qquad
\begin{subfigure}{0.4\textwidth}
\includegraphics[width=0.9\linewidth]{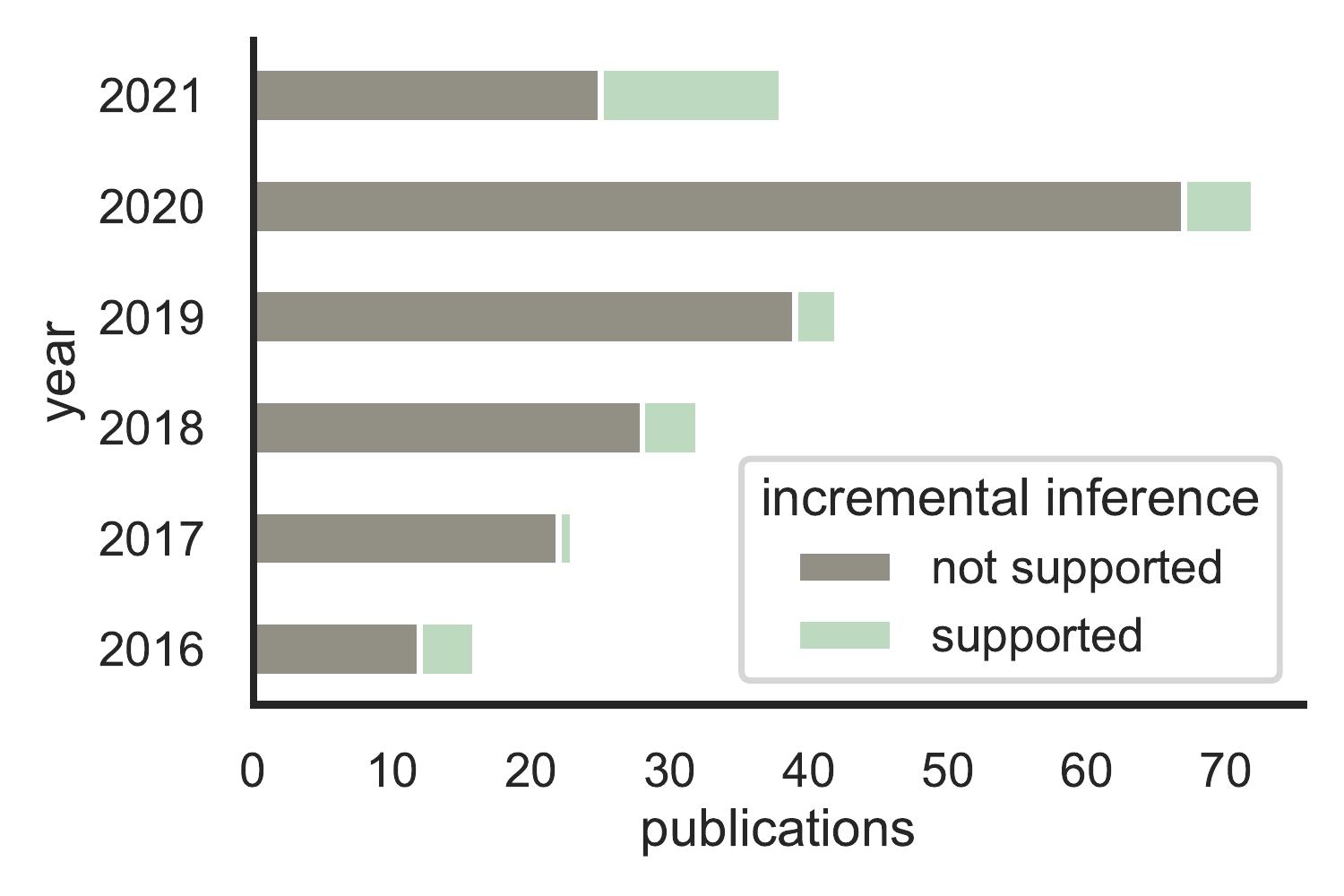}
\caption{Incremental inference.}
\label{fig:inference}
\end{subfigure}

\caption{Support for incremental re-training and inference split by year. Only few approaches can be re-trained or inferred in an incremental fashion.}

\end{figure}


\subsection{Incremental updates}
\cref{fig:retraining} 
shows the number of approaches supporting incremental re-training without major modifications. Disappointingly, only a handful approaches offer support for incremental re-training. This is not surprising since as the last subsection pointed out, most approaches support a transductive setting only. A little more encouraging are the numbers shown in \cref{fig:inference} 
-- there is a small but fairly consistent fraction of approaches that support incremental inference. Overall, though, support for any form of incremental updates is low with no obvious upwards trend during the past six years.

\subsection{Summary of findings}
In summary, our literature review uncovered the following major tendencies in published work from the last six years:
\begin{enumerate}
    \item Most work focused on the transductive setting, and almost all approaches were evaluated in a transductive setting regardless of the model's capabilities. 
    \item There has been some work in recent years on user-inductive models, but it is almost exclusively focused on sequential modeling such as session-based recommendations.
    \item The overwhelming majority of approaches does not support incremental updates.
\end{enumerate}
These findings demonstrate clearly how underexplored inductive generalization scenarios are and show an obvious lack of models that can be updated incrementally with new data. 

\section{Case studies}
\label{sec:case_studies}
To complement our findings above and illustrate key limitations, we now review techniques from three broad classes of recommendation algorithms in more depth. \cref{table:case-study} provides a summary of properties of the algorithms discussed below for reference. 

\begin{table*}[ht]
    \begin{center}\small
    \caption{A summary of case studies on representative recommendation methods.} \label{table:case-study}
    \begin{tabular}{ccccccc}
    \toprule
        \textbf{Support for} & \textbf{MF} \cite{koren2009matrix} & \textbf{NeuMF} \cite{he2017neural} & \textbf{itemKNN} \cite{sarwar2001item} & \textbf{GRU2Rec} \cite{hidasi2015session} & \textbf{LightGCN} \cite{he2020lightgcn} & \textbf{IGMC} \cite{zhang2019inductive} \\
         \midrule 
        User-Inductiveness  & \xmark & \xmark & \cmark & \cmark & \xmark & \cmark \\
        Item-Inductiveness  & \xmark & \xmark & \cmark & \xmark & \xmark & \cmark \\
        Incremental Inference & - & - & \xmark & \xmark & - & \xmark \\
        Incremental Re-training & \xmark & \xmark & - & \xmark & \xmark & \xmark \\
         \bottomrule
    \end{tabular}
    \end{center}
    \vspace{-5mm}
\end{table*}

\subsection{Explicit embedding-based approaches}
Embedding-based recommendation techniques are a popular class of methods which seeks to learn user representations ($\bm{e}_u$) and item representations ($\bm{e}_i$) in a joint low-dimensional factor space and leverage different operators (e.g., the inner product operator) between $\bm{e}_u$ and $\bm{e}_i$~\cite{mnih2007probabilistic,hu2008collaborative,pan2008one,koren2009matrix,rendle2009bpr,he2017neural}. Here we review two popular methods, Matrix Factorization (\textbf{MF})~\cite{koren2009matrix} and Neural Collaborative Filtering (\textbf{NeuMF})~\cite{he2017neural}, and discuss their support for the two properties described in \cref{sec:definition}.

\subsubsection{Matrix factorization} In vanilla \textbf{MF}, we seek to learn the above $K$-dimensional user and item embeddings ($\bm{e}_u$, $\bm{e}_i$) and use the inner product to model a user's preference utility on an item, i.e.,
\begin{equation}
    r_{u,i} = \left<\bm{e}_u, \bm{e}_i\right>, \label{eq:mf}
\end{equation}
where $\left<\cdot,\cdot\right>$ denotes the inner product operator.

As the parameter space of the above factorization model relies explicitly on user and item identifiers, we need to know all user and item contexts at training time (i.e., 
$\mathcal{I}_{\mathit{test}}=\mathcal{I}_{\mathit{train}}$ and $\mathcal{U}_{\mathit{test}}=\mathcal{U}_{\mathit{train}}$).
Therefore, the line of \textbf{MF}-based approaches typically falls into \textit{transductive learning}. 

Vanilla \textbf{MF} learns a static representation for items ($\bm{e}_i$) and users ($\bm{e}_u$) and does not perform any dynamic inference at test time. Therefore, in order to accommodate new information, model re-training is necessary. Even though closed-form solutions may exist under special assumptions (e.g., either item or user embeddings are assumed to be fixed), one typically needs to go through the entire dataset for re-training, i.e., \textbf{MF} is \textit{non-incremental}. 


\subsubsection{Neural collaborative filtering}
\textbf{NeuMF} augments vanilla \textbf{MF} to allow for non-linearities between the user-item representations. \textbf{NeuMF} represents user-item scores by the sum of two subscores resulting from applying a Generalized Matrix Factorization (GMF) operator and an Multi-Layer Perceptron (MLP) operator on top of user-item embeddings, i.e.,
\begin{equation}
r_{u,i} = \left<\bm{w}^{(\mathit{gmf})},~ \bm{\phi}^{(\mathit{gmf})}_{u,i}\right> + \left<\bm{w}^{(\mathit{mlp})},~ \bm{\phi}^{(\mathit{mlp})}_{u,i}\right>,
\end{equation}
and
\begin{equation}
\begin{aligned}
        \bm{\phi}^{(\mathit{gmf})}_{u,i} = \bm{e}_u^{(\mathit{gmf})} \circ \bm{e}_i^{(\mathit{gmf})},\quad \quad
    \bm{\phi}^{\mathit{(mlp)}}_{u,i} = \text{\textbf{MLP}}([\bm{e}_u^{(\mathit{mlp})}; \bm{e}_i^{(\mathit{mlp})}]).
\end{aligned}
\end{equation}
Here $\{\bm{e}_u^{(\mathit{gmf})}, \bm{e}_i^{(\mathit{gmf})}\}$, $\{\bm{e}_u^{(\mathit{mlp})}, \bm{e}_i^{(\mathit{mlp})}\}$ are user-item embeddings in the GMF and MLP components respectively, $\bm{\phi}^{(\mathit{gmf})}_{u,i}$ represents the element-wise product between user and item embeddings, and $\bm{\phi}^{(\mathit{mlp})}_{u,i}$ represents the output from the top layer in a learnable MLP component applied on the concatenation of user and item embedings.

Similar to \textbf{MF}, \textbf{NeuMF} requires explicit user and item identifiers when learning embeddings. Hence, neither user-inductive learning nor item-inductive learning is supported by \textbf{NeuMF}. In the same vein, since \textbf{NeuMF} learns explicit embeddings, model re-training is required to enable updated predictions on new data. As this process still requires scanning the training set $\mathcal{D}_{\mathit{train}}$, the original form of \textbf{NeuMF} does not afford incremental updates either.

\subsection{Aggregation-based approaches}
Another line of work represents users through aggregations over their interacted items or \textit{vice versa}. These aggregators can be as simple as a weighted summation with a pre-defined weighting function (e.g., \textbf{itemKNN}~\cite{sarwar2001item}) or a deep neural network which considers sophisticated sequential and temporal interaction dynamics (e.g., \textbf{GRU2Rec}~\cite{hidasi2015session, hidasi2018recurrent}).

\subsubsection{Memory-based collaborative filtering}
Classic memory-based collaborative filtering includes user-centric~\cite{herlocker1999algorithmic} and item-centric~\cite{sarwar2001item,linden2003amazon} methods. As an item-centric approach, \textbf{itemKNN}~\cite{sarwar2001item} represents the item context $\bm{x_i}$ using users' historical interactions on this item. Specifically, it generates a user's preference score for an item $i$ by aggregating the user's past interactions with similar items to evaluate users' common interests on both items, i.e.,
\begin{equation}
    r_{u,i} = \frac{1}{\sum_{j\in\mathcal{N}_{i}} \mathrm{sim}(\bm{x_i}, \bm{x_j})} \sum_{j\in\mathcal{N}_{i}} \mathrm{sim}(\bm{x_i}, \bm{x_j})
    \cdot r^*_{u,j}, \label{eq:itemKNN}
\end{equation}
where $\mathrm{sim}(\cdot,\cdot)$ is a pre-defined similarity function (e.g., cosine or Jaccard similarity) that uses item context vectors $\bm{x_i}$, $\bm{x_j}$ as inputs. $\mathcal{N}_{i}$ denotes a set of items which are similar to the given item $i$.

\textbf{itemKNN} is \textit{fully inductive}. For new test users, their historical item interactions can be added into the user context $\bm{x}_u$ and before recomputing recommendations. Also, for unseen items at test time (i.e., $\bm{x}_i$ is known), the above aggregator is able to generate user preference scores on these items. \textbf{itemKNN} does not require a training step (sometimes called \say{lazy learning}) and hence our definition of incremental re-training 
does not apply. Accommodating new data points can be easily achieved by adding every new data point into item and user contexts ($\bm{x}_i$ and $\bm{x}_u$). 
These changes in item and user contexts, however, generally require re-computing item-item similarities $\mathrm{sim}(\bm{x}_i, \bm{x}_j)$ and updating $r_{u,j}^*$, thus the computation cost of \cref{eq:itemKNN} depends on the size of test data. Therefore, we argue \textbf{itemKNN} does not support incremental updates when implemented off-the-shelf. 




\subsubsection{Deep sequential recommendation}
Sequential or session-based recommendation approaches leverage the sequential (or temporal) dynamics in a user's previously interacted item sequence to predict the user's next action. These approaches typically seek to learn an aggregator over item sequences~\cite{hidasi2015session,tang2018personalized,kang2018self,sun2019bert4rec}. 
For example, \textbf{GRU2Rec} \cite{hidasi2015session} uses a GRU-based RNN for item interactions $i^{(0)}, \ldots, i^{(t-1)}$ through a GRU-based RNN,
\begin{equation}
    \bm{r}_u^{(t)}
    = \text{\textbf{GRU}}( \bm{r}_u^{(t-1)},  \bm{x}_{i^{(t-1)}}),  \\
    \label{eq:gru2rec}
\end{equation}
where $\bm{r}_u^{(t)}$ 
is an $|\mathcal{I}|$-dimensional vector reflecting $u$'s preference utility for each item 
at time $t$, and $\bm{x}_{i^{(t)}}$ denotes the one-hot identifier-based encoding of an item at time $t$. Here \textbf{GRU}$(\cdot)$ refers to the gated recurrent unit that processes item sequence recursively as the input and project the output scores on the entire item set $\mathcal{I}$.

Notice that model parameters inside the GRU module depend on item identifiers, thus the model does not support item-inductive learning. However, since a user's representation only depends on the aggregated items, \cref{eq:gru2rec} can be easily applied on new users thus making it \textit{user-inductive}. 

\textbf{GRU2Rec} does not support incremental re-training as adding new data points to the training data requires fine-tuning the model to fit the entire training set. The original form of \textbf{GRU2Rec} does not afford incremental inference as the GRU aggregator needs to do a forward pass on all previous interacted items hence the computation is related to the length of the test sequence. However if the hidden states of the GRU module are cached, then the inference becomes incremental as the prediction only requires the cached state of the most recent timestamp.

\subsection{GNN-based approaches}
A recent line of work frames user-item interaction data as a graph and applies graph neural networks (GNNs)~\cite{ying2018graph, wang2019neural, he2020lightgcn, zhang2019inductive} to make recommendations. These approaches share their perspective on viewing users and items as nodes and using $r^*_{u,i}$ as edge weights between them. Here we discuss two recent methods, \textbf{LightGCN}~\cite{he2020lightgcn} and  \textbf{IGMC}~\cite{zhang2019inductive}. that take different approaches to learning from graph-based data.

\subsubsection{Graph convolutional networks}
A typical example for the application of a graph convolution network (GCN) to recommendation is \textbf{LightGCN}. \textbf{LightGCN}~\cite{he2020lightgcn} was proposed as an improvement over the more general neural graph collaborative filtering model~\cite{wang2019neural}. On a high level, \textbf{LightGCN} learns user and item identifier-based embeddings and propagates them linearly through the user-item interaction graph. Such a propagation is achieved through a linear convolution operator, i.e.,
\begin{equation}
    \begin{aligned}
    \bm{e}_u^{(k+1)} =& \sum_{i\in\mathcal{N}_u}\frac{1}{\sqrt{|\mathcal{N}_u||\mathcal{N}_i|}} \bm{e}_i^{(k)}, \\
    \bm{e}_i^{(k+1)} =& \sum_{u\in\mathcal{N}_i}\frac{1}{\sqrt{|\mathcal{N}_u||\mathcal{N}_i|}} \bm{e}_u^{(k)}, 
    \end{aligned} \label{eq:lightGCN-agg}
\end{equation}
where $\mathcal{N}_i$ and $\mathcal{N}_u$ denote the neighborhoods of item $i$ and user $u$ on the interaction graph, $\bm{e}_i^{(k)}$ and $\bm{e}_u^{(k)}$ denote the $k$-th layer (hop) of the propagation. The model prediction still follows the conventional paradigm which takes the inner product between user and item embeddings, i.e.,
\begin{equation}
    r_{u,i} = \left<\bm{e}_u, \bm{e}_i\right>, \label{eq:lightGCN-prediction}
\end{equation}
where $\bm{e}_u = \sum_{k=0}^K \alpha_k\bm{e}_u^{(k)}$, $\bm{e}_i = \sum_{k=0}^K \alpha_k\bm{e}_i^{(k)}$ are the final user and item representations combined by their embeddings in each convolution layer. The matrix form of the graph convolution-based propagation process for \textbf{LightGCN} can be formulated as
\begin{equation}
    \bm{E}^{(k+1)} = \bm{D}^{-\frac{1}{2}}\begin{bmatrix} \bm{0} & \bm{R}^*\\ (\bm{R}^*)^T & \bm{0} \end{bmatrix}\bm{D}^{-\frac{1}{2}} \bm{E}^{(k)}, \label{eq:lightGCN-matrix}
\end{equation}
where $\bm{R}^*$ is a $|\mathcal{U}|\times|\mathcal{I}|$ matrix to represent the observed interaction user-item graph, $\bm{E}^{(k)}$ denotes the $k$-th layer user and item embeddings, and $\bm{D}$ is a diagonal matrix where the entry represents the degree of each user or item node.

As showed in \cref{eq:lightGCN-matrix}, the parameter space of \textbf{LightGCN} explicitly relies on item and user identifiers. Training such a model requires the full graph Laplacian being given (i.e.,  
$\mathcal{I}_{\mathit{test}}=\mathcal{I}_{\mathit{train}}$ and $\mathcal{U}_{\mathit{test}}=\mathcal{U}_{\mathit{train}}$)
thus falling into the \textit{transductive learning} category.

Note the original form of \textbf{LightGCN} does not capture any sequential or temporal dyanmics in the interaction graph; hence a full model update is needed to adapt to new information. Again, the model needs to scan the training set during fine-tuning, and thus does not support incremental updates.

\subsubsection{Subgraph-based methods}
Instead of learning global embeddings for each node of the graph, recent work attempts to develop on-the-fly aggregators on sampled subgraphs around each user-item edge. One recent example is Inductive Graph-Based Matrix Completion (\textbf{IGMC}) \cite{zhang2019inductive},  which (i) uses relative indexes to represent user and item nodes instead of the global identifiers and (ii) operates the graph convolutions on extracted enclosing subgraphs rather than the complete user-item graph. Specifically, for a target user-item pair $(u,i)$, it extracts the $k$-th hop neighborhoods of both nodes through Breath-First Search (BFS) as well as the interaction edges among these neighbors. The relative indexes are used to represent the node type (i.e., user or item) and the distance to the target nodes and to replace the global node identifiers. 
Then graph convolution aggregators (e.g., \cref{eq:lightGCN-matrix}) can be applied on this extracted subgraph to generate the final user and item embeddings and the preference score $r_{u,i}$.

Since \textbf{IGMC} is trained on extracted subgraphs where nodes are represented by their relative indexes, it is capable of generating predictions on user and item nodes which are unseen in the training data. Hence, \textbf{IGMC} is \textit{fully inductive}.

Although it does not explicitly model temporal dynamics, the inductive property enables \textbf{IGMC} to update predictions by directly incorporating new data points in the subgraph structure. However, such an inference process requires scanning the complete subgraph and thus depends on the size of the existing test data. Therefore, we conclude that the vanilla implementation of \textbf{IGMC} does not support incremental inference. Re-training or fine-tuning \textbf{IGMC} with additional data points needs to pass previous training data as well; hence \textbf{IGMC} does not support incremental re-training either.

\section{Best practices and open challenges}
\label{sec:guidance}
As demonstrated in the two previous sections, there exist substantial gaps between current research and the requirements of real-world recommender systems. Here, we discuss potential solutions to narrow this gap, and present best practices and open challenges for future research. Existing approaches to address inductive learning and incremental updates are summarized in \cref{table:model_summary}.

\begin{table}[]
    \centering
    \caption{A summary of existing approaches for inductive learning and incremental updates.}
    \label{table:model_summary}
    \begin{tabular}{lp{0.6\textwidth}l}
    \toprule
      \textbf{Support for}   &  \textbf{Strategy} & \textbf{Examples}  \\
    \midrule

      Inductive Learning  & Replace item (user) embeddings with feature-based representations & \cite{wu2020learning, zhao2020catn, li2020ddtcdr, he2016vbpr, wang2015collaborative} \\
      & Represent users (items) by aggregating over item (user) embeddings & \cite{shi2020beyond, zaheer2017deep, wang2020make, hidasi2015session, hidasi2018recurrent} \\
Incremental Updates & Existing practices for incremental inference
\begin{itemize}
\item Cache intermediate model states
\item Fix the length of user history for inference      \vspace{-0.12in} 
\end{itemize}
 & \cite{hidasi2015session, hidasi2018recurrent,you2019hierarchical,tang2018personalized} \\
      & Existing practices for incremental retraining 
      \begin{itemize}
      \item Re-train shallow MF incrementally \vspace{-0.12in}
      \end{itemize}      
      & \cite{wang2018streaming, zhang2020retrain, guo2019streaming, lin2018online} \\
    \bottomrule
    \end{tabular}
\end{table}

\subsection{Models for inductive learning}
\subsubsection{Possible solutions}
As shown in \cref{sec:case_studies}, 
an important design choice that prevents current recommendation models from being inductive is relying on explicitly instantiated embeddings for each user or item. These latent embeddings have to be trained or fine-tuned, because there is no close-form solution due to the non-convex nature of many state-of-the-art models. Prior literature in cold-start, sequential, and session-based recommendations suggests that this problem can be partially addressed in the following two ways.

The first way is to replace the explicit embeddings with feature-based representations~\cite{wu2020learning, zhao2020catn, li2020ddtcdr}:
\begin{equation}
    \bm{e}_{u} = \phi(\bm{x'}_{u}), \qquad \bm{e}_{i} = \psi(\bm{x'}_{i})
\end{equation}
where $\bm{x'}_{u}, \bm{x'}_{i}$ are feature vectors of users (items) but do not contain explicit identifiers and $\phi(\cdot), \psi(\cdot)$ are mapping functions. Feature vectors of users can for example be constructed from users' demographic information or previously elicited preferences. Item features could be extracted through encoders such as CNNs for images~\cite{he2016vbpr}, or autoencoders for text~\cite{wang2015collaborative}. 

Second, one can use composed representations that represent users (items) by aggregating over item (user) embeddings. For example, to make recommenders user-inductive, one can construct user representation $\bm{e}_u$ by
\begin{equation}
    \bm{e}_u = g(\{\bm{e}_i|i \in \bm{x}_u\})
\end{equation}
where $g$ is an aggregation functions over interacted items ($i \in \bm{x}_u$) so that during inference time, embeddings for unseen users can be generated with a direct forward pass. Common aggregation functions include mean or max pooling, set (permutation invariant) aggregators~\cite{shi2020beyond, zaheer2017deep}, and sequence aggregators~\cite{wang2020make, hidasi2015session, hidasi2018recurrent}. Item-inductive recommenders can be achieved similarly.

\subsubsection{Open challenges}
Even though feature-based approaches offer a possible way to make a model inductive, there are fundamental limitations of such approaches. The most obvious limitation is that features are not always available in practice or cannot be accessed to due privacy reasons. Furthermore, features may fall short in representing diverse and latent aspects of user preferences~\cite{knijnenburg2011each}.
Deriving good features requires substantial domain expertise; in many domains it is not even clear how semantically meaningful features should look like, e.g., art. 
This brings important research questions for future work: \textit{Can we combine feature-based information with explicit embeddings and still maintain an inductive model?} \textit{If so, what are the trade-offs and how can be optimize them?} 
Moreover, future work should investigate these questions with a dynamic environment in mind, as explicit features are typically static and cannot be readily updated as implicit representations.  

Aggregation-based methods circumvent the issues above, but suffer from the the \emph{curse of the directionality.} Current aggregation-based methods have to make a choice and can only represent either items through users or vice-versa~\cite{shi2020beyond, hidasi2015session}. This is fundamentally due to the directionality of the aggregation in the forward-pass as one has to serve as the inputs for the other. The fundamental quest here is: \textit{Can we design models that can form cyclic aggregation relationships between users and items for a fully inductive model?} 
Recently developed methods, e.g., subgraph-based strategies~\cite{zhang2019inductive} and permutation equivariant operators~\cite{hartford2018deep}, are promising directions and can serve as good starting points. 

The fact that latent factors typically require a gradient back-propagation process poses fundamental challenges to inductive learning. Future work should explore broader model families, such as non-parametric and tree-based models, to enable inductive recommendations. A bigger related question is also how properties of the data (sparsity, domain size, etc.) impact algorithm performance~\cite{deldjoo2021explaining}.
    
\subsection{Model support for incremental updates}
\subsubsection{Possible solutions.}
There are three larger groups of techniques that past and current research and industry take to enable incremental model updates. 

The first group of techniques relies on caching intermediate model states. For sequential models, especially those based on RNNs and their close variants, practitioners often cache the latest hidden states for each user $\bm{r}_u^{(t-1)}$, so that incremental inference only requires a one-step forward pass~\cite{hidasi2015session, hidasi2018recurrent} (Eq.~\ref{eq:gru2rec}). However, this trick is only applicable to a small number of model classes and fails whenever attention or self-attention mechanisms are present.

Another operation commonly used in practice is to only look back a fixed length $L$ of user history for incremental inference so that the computational cost stays constant, i.e., only processing $i^{(t-L)}, \ldots, i^{(t-1)}$ in Eq.~\ref{eq:gru2rec}. However, this approach is only applicable to user-inductive models and can by design only memorize very short-horizon user preferences.

The last group of techniques concerns the re-training of shallow MF models. Prior work showed that shallow MF models can be re-trained incrementally by sampling a fixed-size data from the training set~\cite{wang2018streaming, guo2019streaming} or meta-learning~\cite{zhang2020retrain}. 
However, it is unclear if and how incremental re-training can be designed for more complex models.

\subsubsection{Open challenges}
The solution approaches discussed above are mostly ad hoc, limited in their scope, and can impact model performance substantially. This points out fundamental research challenges for future work.

First, we need models that support incremental inference natively. As a necessary step towards this goal, we need to explore constructs and conceptual models beyond those of matrix reconstruction, such as streams, trees, graphs, etc. 
    
Another challenge concerns the re-training of deep neural networks. Compared to shallow MF-based methods, deep models are non-trivial to re-train incrementally because of their large parameter spaces and highly non-linear cost functions. One approach to enabling incremental re-training is to approximate the original objective through functions that can be quickly optimized. This would be a generalization of \emph{fold-in} methods for MF-based models where the original objective is simplified by treating either item or user embeddings as fixed. Other approaches include meta-learning and fine-tuning which have shown great promise in other areas ~\cite{devlin2019bert,yu2020meta} and should be further explored in future work.

When compared to transductive or user-inductive models, enabling incremental updates for item-inductive or fully inductive models presents additional challenges as the new data brings in unseen items. Most of the techniques discussed above will not work as they generally assume that the item base does not change over time~\cite{wang2018streaming, zhang2020retrain, hidasi2015session, hidasi2018recurrent}. For latent factor-based models, this implies that embeddings for new items need to be added and trained in real time. Future work should explore model design that allows flexible changes of underlying item set during incremental updates.

Real world user-item interactions do not generally follow the i.i.d. data assumption that most recommendation models are built on. For example,  product seasonality can affect people purchasing behavior, and breaking news may skew the distribution of content interactions. Future work should develop methods that can allow incremental updates to continue function even under such non-stationary settings. A potential research direction is to draw the rich literature from online machine learning~\cite{shalev2011online} where non-stationarity is a commonly faced challenge.

A last important challenge is to support model updates for general data editing. So far, our discussions on incremental updates have been focusing on new data points arriving in real-time. However, due to emerging privacy regulations, such as GDPR, it becomes increasingly critical to support scenarios where users want to remove their data records~\cite{wen2018exploring}. In such situations, today's approaches -- even sequential ones -- 
will mostly fail to support such operations. 
Therefore, how to best honor data removal through incremental updates is an emerging issue for future research.

\subsection{Improving evaluation}
\subsubsection{Evaluating inductive models} As our results show in \cref{sec:state-of-research},
most of the recommendation systems developed in the last five years were evaluated in transductive settings. We argue that a transductive setting is highly impracticable for most recommender systems in the real-world. When evaluation is only done in a transductive setting, practitioners may run the risk of over-estimating a models' true performance. Instead, we propose the following base protocol for future offline inductive evaluations, splitting by the specific inductive scenario they are testing:

\begin{itemize}
    \item To conduct \textbf{user-inductive evaluation}, given a dataset $\mathcal{D}$, one should divide $\mathcal{D}$ into $\mathcal{D}_{\mathit{train}}$ and $\mathcal{D}_{\mathit{test}}$ \textit{across users}, that is, there should be no overlap between $\mathcal{U}_{\mathit{train}}$ and $\mathcal{U}_{\mathit{test}}$. Some prior literature refers to this setting as strong generalization~\cite{marlin2004modeling}. 

    \item \textbf{Item-inductive evaluation} can be achieved by dropping all records pertaining to a set of items $\mathcal{I}_\mathit{drop}$ from the training set, that is $\mathcal{D}_{\mathit{train}}^\prime = \mathcal{D}_{\mathit{train}} \setminus \{(\xuu, \xii, r^*_{u,i}) | \xii \in \mathcal{I}_{\mathit{drop}}\}$, which is similar to new cold-start item evaluation employed by prior work~\cite{deldjoo2019movie}. Different ways of choosing $\mathcal{I}_{\mathit{drop}}$ test distinct aspects of a model's generalization ability, e.g., random selection assumes an i.i.d. setting, popularity-based selection shows models' ability to generalize to long-tail items, and attribute-based selection can be leveraged to test cross-attribute transferability.
    
    \item \textbf{Fully inductive evaluation} should follow both steps presented above --- first split the dataset according to user-inductive evaluation and then mask out item records per item-inductive evaluation.
    
\end{itemize}


\begin{algorithm}[!tbp]
  \KwIn{$\mathcal{D}(u)$, randomly initialized model $M_{-1}$}
  \KwOut{recommendation performance $p$, cost of updates $c$.}
   divide $\mathcal{D}(u)$ into $T+1$ disjoint buckets: $\mathcal{D}_\mathit{batch}(u)$, $\mathcal{D}_1(u)$, $\mathcal{D}_2(u)$, ..., $\mathcal{D}_T(u)$\;
  $M_0$ $\leftarrow$ batch training $M_{-1}$ using $\mathcal{D}_\mathit{batch}(u)$\;
  initialize recommendation performance $p$\;
  initialize cost of updates $c$\;
  \For{$t$ from $1$ to $T-1$}{
    update $M_{t-1}$ to $M_{t}$ with $\mathcal{D}_t(u)$ and return cost $c_t$\;
    $p_t$ $\leftarrow$ evaluate $M_t$ on $\mathcal{D}_{t+1}(u), ..., \mathcal{D}_T(u)$\;
    append $c_t$ to $c$ and $p_t$ to $p$\;
  }
  \caption{Proposed evaluation protocol for incremental updates. The cost of the updates should be explicitly measured alongside ranking performance.}
  \label{algo:incremental_eval}
\end{algorithm}

\subsubsection{Evaluating incremental updates}
The static ``hold-one-out'' or ``hold-x\%-out'' protocol commonly adopted by existing work fails to evaluate models' incremental updates performance. We propose that future work should evaluate recommenders through a replay protocol (\cref{algo:incremental_eval}),
similar to how cold start performance or a contextual bandit is evaluated with two important differences. First, when making updates, the algorithm has to use the parameters from the previous round; second, updates need to be efficient which is why computational cost is also tracked.
The basic idea of \cref{algo:incremental_eval} is to closely mimic the data and model serving flow of real production systems. At each time point $t$, a model needs to incorporate new data points coming in and then make predictions about future user behavior. 
Beyond the immediate next bucket ($\mathcal{D}_{t+1}(u)$), one can also look the results from multiple ($K$) steps ahead (i.e., $\mathcal{D}_{t+1}(u), ..., \mathcal{D}_{t+K}(u)$) so that 
recommenders are encouraged to promote long-term user satisfaction and success. 
It is encouraging to see that recent work~\cite{jugovac2018streamingrec} starting formalize the replay protocol in their open-source library. However, we want to emphasize that for incremental recommendation, evaluating the cost of updates is as important as evaluating the ranking performance which is why it is also explicitly mentioned in~\cref{algo:incremental_eval}. Another metric to consider is the gap between the incrementally updated model and the \emph{skyline performance} of a model that has been fully re-trained at various checkpoints. 

\subsubsection{Connections to RL and online learning.} When moving to a sequential evaluation scheme, there are several issues that need special consideration. First, a critical issue even in traditional transductive recommendation is the fact that we do not observe ratings on all items, and rating data is missing not at random leading to severe biases~\cite{schnabel2016}. This stems from the fact that in the language of reinforcement learning a recommender system (including its interface and other components), is actually a policy $\pi(s)$. This policy is mapping that given state $s$ as input, computes a set of items to users, and only gets to observe feedback on some or all items that were shown. When we want to evaluate a new recommender systems, i.e., a new policy $\pi'(s)$, we need to be able to reliably answer how relevant unrated recommendations are. In most of past work, unrated items are implicitly assumed to be irrelevant, biasing the evaluation heavily towards the policy $\pi(s)$ under which the recommendations were collected. Except from trying to collect fully rated datasets~\cite{kuairec}, there are roughly two groups of approaches to deal with this. In off-policy evaluation, observed data samples are re-weighted to compensate for selection biases~\cite{swaminathan2015counterfactual}, but this may result in high variance estimates~\cite{saito2020open} and also requires us to have some randomness in our deployed policy~\cite{li2011unbiased}. Another option is to specify or model all user responses via a user simulator~\cite{ie2019recsim,zhang2020evaluating}. However, the latter runs the risk of being circular -- it is the user responses which  we are out to estimate in the first place. 

The second issue is that in our discussion so far, we have assumed that users' preferences are constant throughout the process. This also means that recommendations at each time step do not impact future preferences. This may be a reasonable assumption for some domains where we expect long-term interests to dominate -- say in movie recommendation -- but less credible in other domains, e.g., fashion. In reinforcement learning, this complexity is modeled via the state $s$ of a user. This not only allows us to capture changing preferences, but also define metrics (rewards) that depend on the sequence of things previously shown.   

Finally, in order to get a clearer picture of algorithm performance, there is the question of how to quantify the performance gap between how well $M_t$ doing in each round $t$ with the data available vs. how well an oracle model (knowing all data) could have done. In online learning, a model gets to make a decision in each round $t$ after observing new data and then gets to observe a reward (or loss). To compare algorithms, a common notion is that of \emph{static regret}, where we measure the gap between how well the best model on all data would have done vs. how well we did in each time step~\cite{shalev2011online}. However, this definition does not work for non-stationary environments where the best preference model can change from timestep to timestep. 



\subsubsection{Open challenges}
Similar to model design, the evaluation protocols that we proposed above mostly focus on recommendations with stationary data streams (data coming from an i.i.d. distribution). Evaluating inductive learning and incremental updates under non-stationary settings remains an open challenge as models may be drift over time. Future work could explore the use of notions like \emph{dynamic regret}~\cite{zinkevich2003online} from online learning, where performance is measured relative to a dynamic reference set. An even more flexible approach would be to use a full user simulator where dynamic recommendation policies evaluated~\cite{ie2019recsim}. Another open question is how robust evaluation protocols are when certain assumptions are violated in the data (e.g., data missing at random, stationarity).

\section{Conclusion}

In this paper, we argued that inductive learning and incremental updates are two critical properties of practical recommender systems and provided formal definitions for future research. We investigated the current state of recommendation research by extensively reviewing related papers published from 2016 to 2021, and conducting case studies on several representative recommendation methods. Our study revealed a concerning trend that current research literature is largely focused on the \textit{transductive} setting and also mostly does not support incremental updates. This is far from the practical need for flexible, efficient, and fresh recommendations in production. We synthesized some existing ad hoc solutions that partially address the these needs, but the general problem of supporting unseen users and items and incrementally updating recommendations remain unsolved. We hope this survey paper can help drive more attention to these important and under-explored areas and our proposed research roadmap can help guide and inform future investments.

\bibliographystyle{ACM-Reference-Format}
\bibliography{sample-base}


\end{document}